\documentclass{article}
\usepackage{amsfonts,amssymb, amsmath}
\usepackage[english]{babel}
\usepackage{graphicx}

\def\g{{\gamma}}

\def\nn{\nonumber }
\def\bq{ \begin{equation}}
\def\eq{ \end{equation}}
\def\ben{ \begin{eqnarray}}
\def\en{ \end{eqnarray}}

\newtheorem{prop}{Proposition}

\newtheorem{defi}{Definition}

\newtheorem{re}{Remark}
\newenvironment{rem}{\begin{re} \rm }{\end{re}}

\newtheorem{exa}{Example}

\begin{document}

\title{Integrable  discretization and deformation of the nonholonomic Chaplygin ball}
\author{A.V. Tsiganov \\
\it\small St.Petersburg State University, St.Petersburg, Russia\\
\it\small e-mail:  andrey.tsiganov@gmail.com}

\date{}
\maketitle

\begin{abstract}
The rolling of a dynamically balanced ball on a horizontal rough table without slipping  was  described by Chaplygin
using  Abel quadratures. We discuss integrable discretizations and deformations of this nonholonomic system using the same Abel quadratures. As a  by-product one gets new geodesic flow on the unit two-dimensional sphere whose additional integrals of motion are  polynomials in the momenta of fourth order.
 \end{abstract}

\section{Introduction}
\setcounter{equation}{0}
The nonholonomic Chaplygin ball  is that of a dynamically balanced three-dimensional ball that rolls on a horizontal table without slipping or sliding \cite{ch03}.  'Dynamically balanced' means that the geometric center coincides with the center of mass but
the mass distribution is not assumed to be homogeneous.  Because of the roughness of the table this ball cannot slip, but it can turn about the vertical axis without violating the constraints. There is a large body of literature dedicated to the Chaplygin ball, including the study of its generalizations, see  \cite{bm01,bm08, bmb13,bm14,dus04,fed07,ts11} and references within.

In \cite{ts17v} we discuss integrable  discretizations and deformations of the nonholonomic Veselova system using standard divisor arithmetic on the hyperelliptic curve of genus two. Because   nonholonomic Veselova system is equivalent to the nonholonomic Chaplygin ball with a pure mathematical point of view \cite{ts12,ts12b}, we can transfer obtained results on the Chaplygin case.

The main our aim is to describe integrable  discretizations and deformations of the nonholonomic Chapligin ball
in a such way that we can eliminate the corresponding nonholonomic constraint and turn back to the holonomic Euler top on any step of our construction. In addition, we discuss auto and hetero B\"{a}cklund transformations  for the free motion on the sphere (partial case of Euler top) associated with the arithmetic of divisors on an auxiliary plane curve. As a result, we obtain the new geodesic flow on the sphere with the quartic integral of motion.

 The paper is organized as follows.  The second Section  is devoted to well-studied description of the Chaplygin ball motion in term of Abel's quadratures.   Section 3 deals with the particular case when the angular momentum vector is parallel to the plane.  Using Chaplygin's calculations, we will introduce Abel's differential equations and an intersection divisor. Then we will briefly  repeat main information about the arithmetic of divisors and will explicitly  show one of the possible discretizations  of the Chaplygin ball motion associated with  this arithmetic. In Section 4 we will present  new integrable  nonholonomic and holonomic systems, which can be considered as integrable deformations of the Chaplygin ball and Euler top restricted to the sphere.

\section{Chaplygin ball}
Following \cite{ch03} consider a dynamically non-symmetric ball rolling without slipping over a horizontal plane.
Its mass, inertia tensor and radius will be denoted by $m$, $\mathbf I = \mathrm{diag}(I_1, I_2, I_3 )$ and $r$ respectively.
Assume also that the mass center and the geometric center of the sphere coincide with each other.

Let $\g=(\g_1,\g_2,\g_3)$ is the vertical unit vector, $\omega=(\omega_1,\omega_2,\omega_3)$ and  $v=(v_1,v_2,v_3)$ are respectively the angular velocity of the ball  and the velocity of its center.  The condition of non-slipping of the point  of
contact of the ball with the horizontal plane is
\bq\label{nh-con}
v+r\,\omega\times \g=0\,.
\eq
Here and below $\times$ denotes the vector product in $\mathbb R^3$ and all the vectors are expressed in the so-called body frame, which is firmly attached to the ball, and its axes coincide with the principal inertia axes of the ball.

Under the nonholonomic constraint (\ref{nh-con}) the equations of motion can be reduced to the following closed system of
equations of motion
\bq\label{m-eq}
  \dot \g=\g\times \omega\,, \qquad \dot M=M\times \omega\,,
\eq
which have the same form as the Euler-Poisson equations in rigid body dynamics \cite{ch03}.
Here  $M=(M_1,M_2,M_3)$ is the angular momentum of the ball with respect to the contact point
\bq\label{omega}
 \omega=\mathbf A_\g M\,,
\eq
where in the Chaplygin case
\bq\label{ch-sol}
 \mathbf A_\g=\mathbf A +d\mathrm g^{2}\,\mathbf A\,\g \otimes\g\,\mathbf A\,,\qquad
 \mathbf A=\left(
 \begin{array}{ccc}
 a_1 & 0 & 0 \\
 0 & a_2 & 0 \\
 0 & 0 & a_3
 \end{array}
 \right),\qquad a_k=(I_k+d)^{-1}\,,
\eq
and
\bq\label{g-fun}
\mathrm g=\frac{1}{\sqrt{1-d(\g,\mathbf A\g)}}\equiv
\frac{1}{\sqrt{1-d(a_1\g_1^2+a_2\g_2^2+a_3\g_3^2)}}
\,,\qquad\qquad d=mr^2\,.
\eq
According to \cite{bm01} the system (\ref{m-eq})  determines conformally Hamiltonian vector field
\[
X=\mathrm g\, \Pi\, \mathrm dH\,,\qquad  H=\dfrac12\,(M,\omega)
\]
with respect to the Poisson bivector
\bq\label{chap-poi}
\Pi=\mathrm g^{-1}\,\left(\begin{array}{cc}0&\mathbf \Gamma\\ \Gamma &
\mathbf M\end{array}\right)-d\mathrm g\,(M,\mathbf A\g)\left(\begin{array}{cc}0&0\\ 0&
\mathbf \Gamma\end{array}\right)\,,
\eq
where
\[
\mathbf \Gamma=\left(
                 \begin{array}{ccc}
                   0 & \g_3 & -\g_2 \\
                   -\g_3 & 0 & \g_1 \\
                   \g_2 & -\g_1 & 0
                 \end{array}
               \right)\,,\qquad
\mathbf M=\left(
                 \begin{array}{ccc}
                   0 & M_3 & -M_2 \\
                   -M_3 & 0 & M_1 \\
                   M_2 & -M_1 & 0
                 \end{array}
               \right)\,.
\]
Vector field $X$ possesses four independent first integrals
\bq\label{chap-int}
C_1=(\g,\g)=1,\qquad C_2=(\g,M),\qquad H_1=(\omega,M)\,,\quad H_2=(M,M)\,.
\eq
Integrals of motion $C_{1,2}$ are the Casimir functions of $\Pi$
\[\Pi\mathrm d C_{1,2}=0\,,\]
which give rise to the trivial vector fields. The two remaining integrals of motion $H_{1,2}$
are in the involution with respect to the corresponding Poisson bracket
\bq\label{chap-br}
\{M_i,M_j\}=\varepsilon_{ijk}\left(\mathrm g^{-1}{M_k}-d\mathrm g (M, \mathbf A x) x_k\right),\quad
 \{M_i, x_j\}=\varepsilon_{ijk}\mathrm g^{-1} x_k\,,\quad
\{ x_i, x_j\}=0.
\eq
Here $\varepsilon_{ijk}$ is a totally skew-symmetric tensor.
\begin{rem}
 At $d=0$ one gets standard Euler-Poisson equations for the Euler-Poinsot top, which is a well-studied  holonomic Hamiltonian dynamical system, and standard Lie-Poisson bracket on Lie algebra $e^*(3)$.
\end{rem}


\subsection{Integration of the equations of motion}
In \cite{ch03}  Chaplygin found linear trajectory isomorphism with between two dynamical systems (\ref{m-eq}) with $(\g,M)= 0$ and with $(\g,M)\neq 0$. Modern discussion of the Chaplygin's transformation may be found in \cite{bkm12,dus04}.

Namely, integral of motion $H_1$ is equal to
\[
H_1=(\omega, M)=(\mathbf A_\g M,M)=({\mathbf A}'_\g M,M)+\mathrm g^2(\g,M)(\g,{\mathbf A}''M)\,,
\]
where ${\mathbf A}'_\g$ and ${\mathbf A}''$ are diagonal matrices
 \[
{\mathbf A}'_\g=\mathbf A+d\mathrm g^2\left(
 \begin{array}{ccc}
 (a_1-a_3)(a_1-a_2)x_1^2&0&0\\
 0&(a_2-a_3)(a_2-a_1)x_2^2&0\\
 0&0&(a_3-a_1)(a_3-a_2)x_3^2\\
 \end{array}\right)
 \]
 and
 \[
{\mathbf A}''_\g =\left(\begin{array}{ccc}
 a_1(a_2+a_3)-a_2a_3&0&0\\
  0&a_2(a_1+a_3)-a_1a_3&0\\
  0&0&a_2(a_1+a_2)-a_1a_2\\
 \end{array}\right)\,.
 \]
  So,  on the symplectic  leaves defined by the following values of the Casimir functions
  \[
  C_1=(\g,\g)=1\,,\qquad C_2=(\g,M)=0
  \]
 we can consider integrals of motion
\[H_1= ({\mathbf A}'_\g M ,M)\,,\qquad H_2=(M,M)\]
with diagonal matrix ${\mathbf A}'_\g$ instead integral $H_1$ with non-diagonal matrix $\mathbf A_\g$ (\ref{ch-sol})  when $(\g,M)\neq 0$.

Following \cite{ch03} we can integrate equations of motion at $(\g,M)=0$ using sphero-conical coordinates $q_1$ and $q_2$, which are  the roots of the equation
\[
\dfrac{\g_1^2}{x-e_1}+\dfrac{\g_2^2}{x-e_2}=\dfrac{\g_3^2}{x-e_3}=0\,,\qquad e_i=\dfrac{d}{I_i}\,.
\]
These variables satisfy the following equations
\bq\label{quad-1}
\begin{array}{c}
\dot{q}_1=\dfrac{q_2^2(1+q_1)(e_1-q_1)(e_2-q_1)(e_3-q_1)\Bigl(h_2q_1-dh_1(1+q_1)\Bigr)}{d^2(q_1-q_2)}\\
\\
\dot{q}_2=\dfrac{q_1^2(1+q_2)(e_1-q_2)(e_2-q_2)(e_3-q_2)\Bigl(h_2q_2-dh_1(1+q_2)\Bigr)}{d^2(q_2-q_1)}\,,
\end{array}
\eq
where $h_{1,2}$ are values of the integrals of motion $H_{1,2}$. These equations can be reduced to the Abel quadratures after suitable  change of time \cite{ch03}. Theta function solution for the Chaplygin system in this, as well as in the generic case, was obtained in \cite{fed09}.

We prefer to use other coordinates $u_{1,2}$
\[
u_1 =\dfrac{q_1}{d(q_1+1)}\,,\qquad u_2 =\dfrac{q_2}{d(q_2+1)}\,,
\]
 which are the roots of the equation
\[
\mathrm g^2\,\left(\dfrac{\g_1^2(1-da_1)}{\lambda-a_1}+\dfrac{\g_2^2(1-da_2)}{\lambda-a_2}+
\dfrac{\g_3^2(1-da_3)}{\lambda-a_3}\right)=0\,.
\]
In contrast with  $q_{1,2}$ these coordinates $u_{1,2}$ remain well-defined when $d=0$ \cite{ts11}.

According \cite{ts11,ts12b},  let us consider Poisson map $(u_1,u_2,p_{u_1},p_{u_2})\to (\g_1,\g_2,\g_3,M_1,M_2,M_3)$:
{\setlength\arraycolsep{1pt}
\ben \label{x-Trans}
\g_i&=&\sqrt{\dfrac{(1-da_j)(1-da_k)}{(1-du_1)(1-du_2)}}
\,\cdot\,\sqrt{\dfrac{(u_1-a_i)(u_2-a_i)}{(a_j-a_i)(a_k-a_i)}}\,
,\qquad i\neq j\neq k\,,\\\
\nn\\
M_i&=&\dfrac{2\varepsilon_{ijk}\g_j\g_k(a_j-a_k)\mathrm g}{u_1-u_2}\,
\Bigl((a_i-u_1)(1-du_1)p_{u_1}-(a_i-u_2)(1-du_2)p_{u_2}\Bigr)\,.
\nn
\en}
which relates canonical Poisson bivector and bracket
\bq\label{can-p}
\Pi=\left(
    \begin{array}{cccc}
      0 & 0 & 1 & 0 \\
      0 & 0 & 0 & 1 \\
      -1 & 0 & 0 & 0 \\
      0 & -1 & 0 & 0 \\
    \end{array}
  \right)
\,,\qquad \{u_i,p_{u_j}\}=\delta_{ij}\,,\quad \{u_1,u_2\}=\{p_{u_1},p_{u_2}\}=0\,,
\eq
with the Poisson bivector $P$ (\ref{chap-poi}) and bracket (\ref{chap-br})  for the Chaplygin ball.

Substituting (\ref{x-Trans}) into integrals of motion (\ref{chap-int}) one gets integrals of motion
\bq\label{chap-ham-u}
\begin{array}{l}
H_1=\frac{ 4u_2(du_1-1)(u_1-a_1)(u_1-a_2)(u_1-a_3)p_{u_1}^2}{u_1-u_2}+\frac{4u_1(du_2-1)(u_2-a_1)(u_2-a_2)(u_2-a_3)p_{u_2}^2}{u_2-u_1}\,,\\
\\
H_2=\frac{ 4(du_1-1)(u_1-a_1)(u_1-a_2)(u_1-a_3)p_{u_1}^2}{u_1-u_2}+\frac{4(du_2-1)(u_2-a_1)(u_2-a_2)(u_2-a_3)p_{u_2}^2}{u_2-u_1}
\end{array}
\eq
and separation relations
\bq\label{chap-sep-rel}
4(du_k-1)(a_1-u_k)(a_2-u_k)(a_3-u_k)p_{u_k}^2+u_kH_2-H_1=0\,,\qquad k=1,2.
\eq
The corresponding equations of motion are equal to
\bq\label{quad-2}
\begin{array}{c}
\dot{u_1}=\dfrac{4\mathrm g u_2(a_1-u_1)(a_2-u_1)(a_3-u_1)(du_1-1)p_{u_1}}{u_1-u_2}\\ \\
\dot{u_2}=\dfrac{4\mathrm g u_1(a_1-u_2)(a_2-u_2)(a_3-u_2)(du_2-1)p_{u_2}}{u_2-u_1}\,.
\end{array}
\eq
After change of time $t\to\tau$
\[
\mathrm d\tau=\mathrm g \mathrm d t\,,\qquad \mathrm g=\sqrt{
\dfrac{(1-du_1)(1-du_2)}{1-da_1)(1-da_2)(1-da_3)}}
\]
equations (\ref{quad-2}) yield Abel quadratures
\bq\label{chap-ab}
\dfrac{\mathrm du_1}{\sqrt{f(u_1)}}+\dfrac{\mathrm du_2}{\sqrt{f(u_2)}}=4\mathrm d \tau\,,\qquad
\dfrac{u_1\mathrm du_1}{\sqrt{f(u_1)}}+\dfrac{u_2\mathrm du_2}{\sqrt{f(u_2)}}=0\,,
\eq
on the genus two  hyperelliptic curve
\bq\label{chap-curve}
X:\qquad y^2=f(x)\,,\qquad f(x)=4(1-dx)(a_1-x)(a_2-x)(a_3-x)(xh_2-h_1)
\eq
defined by separation relations (\ref{chap-sep-rel}).
\begin{rem}
 On the any step here we can put $d=0$ and consider standard Euler top
in contrast with equations  (\ref{quad-1}) and Abel's quadratures in original Chaplygin paper \cite{ch03}, see discussion in \cite{ts11}.
\end{rem}
Summing up, we can study the motion of the Chaplygin ball   when $(\g,M)=0$ and then apply
Chaplygin's transformation in order to describe motion  when $(\g,M)\neq0$ \cite{bkm12,ch03, dus04}. It allows us to consider
 discretization of the Chaplygin ball  for $(\g,M)=0$ and then apply Chaplygin's transformation in order
 to describe it's discretization for $(\g,M)\neq0$.

\section{Discretization of the Chaplygin ball motion at $(\g,M)=0$}
Discretization of the Chaplygin ball motion, which preserves the same first integrals as the continuous model, except the energy,  was obtained in \cite{fed07} in the framework of the formalism of variational integrators (discrete Lagrangian systems). Our intermediate aim is to describe discretizations of the Chaplygin ball motion preserving all the integrals of motion in the framework of the standard arithmetic of divisors on a hyperelliptic curve.

Suppose that transformation of variables
\bq\label{b-trans}
 B:\quad (u_{1},u_2,p_{u_1},p_{u_2})\to (\tilde{u}_{1},\tilde{u}_2,\tilde{p}_{u_1},\tilde{p}_{u_2})
\eq
preserves Hamilton equations (\ref{quad-2}) and the form of Hamiltonians (\ref{chap-ham-u}), i.e. that new variables satisfy to the same Abel quadratures
\bq\label{chap-ab2}
\frac{\mathrm d\tilde{u}_1}{\sqrt{f(\tilde{u}_1)}}+\frac{\mathrm d\tilde{u}_2}{\sqrt{f(\tilde{u}_2)}}=4\mathrm d\tau\,,\qquad \frac{\tilde{u}_1\mathrm d\tilde{u}_1}{\sqrt{f(\tilde{u}_1)}}+\frac{\tilde{u}_2\mathrm d\tilde{u}_2}{\sqrt{f(\tilde{u}_2)}}=0\,.
\eq
If we put
 \bq\label{xy-u}
 \begin{array}{rr}
 x_{1,2}=u_{1,2},\quad &y_{1,2}=4(du_{1,2}-1)(a_1-u_{1,2})(a_2-u_{1,2})(a_3-u_{1,2})p_{u_{1,2}}\,,\\
 \\
x''_{1,2}=\tilde{u}_{1,2},\quad &y''_{1,2}=-4(d\tilde{u}_{1,2}-1)(a_1-\tilde{u}_{1,2})(a_2-\tilde{u}_{1,2})(a_3-\tilde{u}_{1,2})\tilde{p}_{u_{1,2}}
\end{array}
\eq
 into the difference of  (\ref{chap-ab}) and (\ref{chap-ab2}), one gets a system of Abel's differential equations
\begin{equation}\label{ab-eq-g2}
\begin{array}{c}
\omega_1(x_1,y_1)+\omega_1(x_2,y_2)+\omega_1(x''_1,y''_1)+\omega_1(x''_2,y''_2)=0\,,\\ \\
\omega_2(x_1,y_1)+\omega_2(x_2,y_2)+\omega_2(x''_1,y''_1)+\omega_2(x''_2,y''_2)=0\,,
\end{array}
\end{equation}
where $\omega_{1,2}$  are holomorphic differentials on hyperelliptic curve $X$ of genus two
\[
\omega_1(x,y)=\frac{\mathrm dx}{y}\,,\qquad \omega_2(x,y)=\frac{x\mathrm dx}{y}\,.
\]
According \cite{ab} solutions of Abel's equations (\ref{ab-eq-g2}) are points of intersection of the curve $X$
with another curve $Y$ not containing $X$
\[
Y:\qquad y=\mathcal P(x)\,,\qquad   \mathcal P(x)=b_3x^3+b_2x^2+b_1x+b_0\,.
\]
Thus, we have an intersection divisor of $X$ and $Y$ defined by the following four points on the plane
\bq\label{in-div}
P_{1}=\left(x_{1},y_{1}\right),\quad P_{2}=\left(x_{2},y_{2}\right)
\qquad\mbox{and}\qquad P''_{1,2}=\left(x''_{1},y''_{1}\right),\quad
P''_{2}=\left(x''_{2},y''_{2}\right)
\eq
It allows us  to relate  transformations of variables $B$ (\ref{b-trans}) with the standard arithmetic of divisors.

By viewing the new variables $\tilde{u}_{1,2},\tilde{p}_{u_{1,2}}$ as the old one ${u}_{1,2},p_{u_{1,2}}$ but computed at the next time-step, then transformation $B$ becomes a  discretization of continuous Chaplygin ball motion.

\subsection{Arithmetic of divisors}
Let us  give some brief background on arithmetics of divisors  \cite{eh16,har77}.

A prime divisor on a smooth  variety $X$ over a field $k$ is an
irreducible closed subvariety $Z\subset X$ of codimension one, also defined over $k$.
\begin{defi} A divisor is a finite formal linear combination
\[D =\sum_i m_iZ_i,\qquad m_i\in \mathbb Z\,, \]
of prime divisors. The group of divisors on $X$, which is the free group on the prime divisors, is denoted Div$X$.
\end{defi}
The group of divisors Div$X$ is an additive abelian group under the formal addition rule
 \[\sum m_i Z_i+\sum n_i Z_i=\sum (m_i+n_i) Z_i\,.\]
To define  an equivalence relation on divisors we  use the rational functions on $X$.
Function  $f$ is a quotient of two polynomials; they are
each zero only on a finite closed subset of codimension one in $X$, which is therefore the union of finitely many prime divisors. The difference of these two subsets  define  a principal divisor $div f$
 associated with function $f$. The subgroup of Div$X$ consisting of the principal divisors is denoted by Prin$X$.
\begin{defi}
Two divisors $D, D'\in \mbox{Div} X$ are linearly equivalent
\[D\approx D'\]
if their difference $D-D'$ is principal divisor
\[
D-D'=div(f)\equiv 0\quad \mbox{mod Prin}X\,.
\]
\end{defi}
The Picard group of $X$ is the quotient group
\[
\mbox{Pic}X =\dfrac{\mbox{Div}X}{\mbox{Prin}X}=\dfrac{\mbox{Divisors defined over k}}{\mbox{Divisors of functions defined over k}}\,.
\]
For a general (not necessarily smooth) variety X, what we have defined is not the Picard group, but the Weil divisor class group. For an irreducible normal variety $X$, the Picard group is isomorphic to the group of Cartier divisors modulo linear equivalence. For a thorough treatment see \cite{eh16,har77}.

The Picard group is a group of  divisors modulo principal divisors, and the group operation is formal addition modulo the equivalence relations. These group operations define so-called  arithmetic of divisors in Picard group
\bq\label{add-jac}
D+ D'=D''\qquad\mbox{and}\qquad [\ell] D=D''\,,
\eq
where $D,D'$ and $D''$ are divisors, $+$ and $[\ell]$ denote addition and scalar multiplication by an integer, respectively.

\begin{rem}
There are some other equivalence relations on divisors, for instance homological equivalence, numerical equivalence,
 algebraic equivalence or rational equivalence, which are different from the  linear equivalence $D\approx D'$ \cite{eh16}. It allows us to define relations between divisors which are differ from the standard arithmetic relations (\ref{add-jac}).
\end{rem}

Let $X$ be a hyperelliptic curve of genus $g$  defined by equation
\bq\label{h-curve}
 y^2 + h(x)y = f(x),
 \eq
where $f(x)$ is a monic polynomial of degree $2g + 2$ with distinct roots, $h(x)$
is a polynomial with deg$h\leq g$.  Prime divisors are rational point on $X$  denoted $P_i = (x_i, y_i)$,  and $P_\infty$ is a point at infinity.
\begin{defi}
Divisor $D = \sum m_iP_i$, $m_i\in \mathbb Z$ is a formal sum of points on the curve, and degree of  divisor $D$ is the sum $\sum m_i$ of the multiplicities of points in  support of the divisor
\[
\mbox{supp}\left(\sum m_iP_i\right)=\bigcup_{m_i\neq 0} P_i\,.
\]
\end{defi}
The degree is a group homomorphism deg : Div$X \to\mathbb Z$. Its kernel is denoted
by
\[
\mbox{Div}^0\, X=\{D\in \mbox{Div}C:\,\mbox{deg}D=0\}\,.
\]
 Quotient group of $\mbox{Div}X$ by the group of principal divisors Prin$X$ is called the divisor class group or Picard group.
 Restricting to degree zero, we also define $\mbox{Pic}^0 X= \mbox{Div}^0 X/\mbox{Prin}X$. The groups $\mbox{Pic}X$ and $\mbox{Pic}^0 X$ carry essentially the same information on $X$, since we always have
 \[
\mbox{Pic}X/\mbox{Pic}^0 X \cong \mbox{Div}X/\mbox{Div}^0 X\cong\mathbb Z\,.
 \]
 The divisor class group, where the elements are equivalence classes of degree zero divisors on $X$,  is isomorphic to the Jacobian of  $X$. By abuse of notation, a divisor and its class in Pic$X$ will usually be denoted by the same symbol.

In order to describe elements of Jacobian we can use semi-reduced divisors.
\begin{defi}
A  semi-reduced divisor is divisor of the form
\[
D=\sum m_i P_i -\left( \sum m_i \right) P_\infty=E -\left( \sum m_i \right) P_\infty\,,
\]
where $m_i>0$, $P_i\neq -P_j$ for $i\neq j$, no $P_i$ satisfying  $P_i=-P_ i$ appears more than once and $E$ is an effective divisor.
\end{defi}
For each divisor $\tilde{D}\in \mbox{Div}^0X $ there is a semi-reduced divisor $D$ so that $D\approx \tilde{D}$. However
semi-reduced divisors are not unique in their equivalence class. In \cite{ts17} we used this fact in order to get auto B\"{a}cklund transformations associated with an equivalence relations.
\begin{defi}
A semi-reduced divisor $D$ is called reduced if  $\sum m_i\leq g$, i.e. if the sum of multiplicities is no more that genus of curve $C$.  The reduced degree  or weight of reduced divisor $D$ is defined as $w(D)=\sum m_i$.
\end{defi}
This is a consequence of Riemann-Roch theorem for hyperelliptic curves that for each divisor $\tilde{D}\in \mbox{Div}^0X$ there is a unique reduced divisor $D$ so that $D\approx \tilde{D}$.

\subsection{One example of discretization}
Let us consider adding two full degree genus two divisors
\[
 D+D'=D''
 \]
 with respective supports supp$(D)=\{ P_1,P_2\}\cup \{P_\infty\}$ and  supp$(D')=\{ P'_1,P'_2\}\cup \{P_\infty\}$ such that no $P_i$ has the same $x$-coordinate as $P'_j$. By definition cubic polynomial
 \[\mathcal P(x)=b_3x^3+b_2x^2+b_1x+b_0\]
  interpolates four points $P_1$, $P_2$, $P'_1$, $P'_2$ and, therefore, it has the standard form
 \bq
 \begin{array}{lcl}
 \mathcal  P(x)
&=&\dfrac{(x-x'_2) (x-x'_1) (x-x_2 ) y_1}{(x_1-x'_1 ) (x_1-x'_2) (x_1-x_2)}
+
\dfrac{(x-x'_2) (x-x'_1) (x-x_1 ) y_2}{(x_2-x'_1 ) (x_2-x'_2) (x_1-x_2)}\\
\\
&+&
\dfrac{(x-x'_2) (x-x_2 ) (x-x_1 ) y'_1}{(x'_1-x_1 ) (x'_1-x_2 ) (x'_2-x'_1)}
+
\dfrac{(x-x'_1) (x-x_2 ) (x-x_1 ) y'_2}{ (x'_2-x_1 ) (x'_2-x_2 )(x'_2-x'_1)}\,,
 \end{array}
 \eq
due to Lagrange interpolation.  Substituting $y=\mathcal P(x)$  into the definition of  genus two hyperelliptic curve $X$
 \[
 y^2=f(x)\,,\qquad f(x)=a_6x^6+ a_5x^5+ a_4x^4+ a_3x^3+ a_2x^2+a_1x+a_0\,.
 \]
we obtain the so-called Abel polynomial \cite{ab}
 \[
\psi(x)=\mathcal P(x)^2-f(x)=(b_3^2-a_6)(x-x_1)(x-x_2)(x-x'_1)(x-x'_2)(x-x''_1)(x-x''_2)\,.
\]
 Equating coefficients of $\psi$   gives abscissas of points $P_1''$ and $P''_2$:
\ben
x''_1+x''_2&=& -x_1-x_2-x'_1-x'_2+\frac{a_5-2b_2b_3}{b_3^2-a_6}\,, \label{g2-x}\\
\nn\\
x''_1x''_2&=&\frac{2b_1b_3+b_2^2-a_4}{b_3^2-a_6}
-(x_1+x_2+x'_1+x'_2)(x''_1+x''_2)
-x_1(x_2+x'_1+x'_2)\nn\\
\nn\\
&-&x_2(x'_1+x'_2)-x'_1x'_2\,.\nn
\en
whereas the corresponding ordinates $y''_{1,2}$  are equal to
\bq\label{g2-y}
y''_{3,4}=-\mathcal P(x''_{1,2})\,.
\eq
Using explicit formulae for the corresponding  transformation $B$  (\ref{b-trans})  we can prove the following statement.
  \begin{prop}
Equations (\ref{xy-u}) and (\ref{g2-x}-\ref{g2-y})   determine transformation $B: (u,p_u)\to (\tilde{u},\tilde{p}_u)$ preserving the form of the canonical Poisson bracket (\ref{can-p}), i.e.
\[
\{\tilde{u}_i,\tilde{p}_{u_j}\}=\delta_{i,j}\,,\qquad \{\tilde{u}_1,\tilde{u}_2\}=\{\tilde{p}_{u_1},\tilde{p}_{u_2}\}=0.
\]
This canonical transformation also  preserves the form of integrals of motion  (\ref{chap-ham-u}).
\end{prop}
The proof is a straightforward calculation.

We can rewrite addition of divisors (\ref{g2-x}-\ref{g2-y}) as canonical transformation of initial  variables  $(\g,M)\to (\tilde{g},\tilde{M})$ using (\ref{x-Trans})  and any modern  computer algebra system. We  do not show these bulky expressions here because the main our aim is the construction of new integrable systems instead of explicit construction of possible  discretizations of the Chaplygin ball motion.

\begin{rem}
Construction of  integrable discretizations associated with doubling of divisors and other arithmetic operations in Jacobian is discussed in \cite{ts17,ts17d,ts17v}.
\end{rem}

\section{Integrable deformation of the Chaplygin ball at $(\g,M)=0$}
According \cite{ts15a,ts15b,ts15c}  we can apply hidden symmetries  of the generic level set of integrals of motion to construct new canonical variables and new new integrable systems  on the initial phase space. Indeed, let us consider  transformations $B$  (\ref{g2-x}-\ref{g2-y}) when $P'_1=P_\infty$ and $P'_2=(d^{-1},0)$ is the ramification point. In this case $B$ has the following form in original variables
 \bq\label{chap-keyless}
 \begin{array}{l}
\tilde{\g}_1^2=\dfrac{1}{(a_1d-1)\tilde{\mathrm g}^2}\left(\g_1^2-\dfrac{(\g_1^2+\g_2^2+\g_3^2)(M_2^2+M_3^2)}{M_1^2+M_2^2+M_3^2}\right)\,,\\
\\
\tilde{\g}_2^2=\dfrac{1}{(a_2d-1)\tilde{\mathrm g}^2}\left(\g_2^2-\dfrac{(\g_1^2+\g_2^2+\g_3^2)(M_1^2+M_3^2)}{M_1^2+M_2^2+M_3^2}\right)\,,\\
\\
\tilde{\g}_3^2=\dfrac{1}{(a_3d-1)\tilde{\mathrm g}^2}\left(\g_3^2-\dfrac{(\g_1^2+\g_2^2+\g_3^2)(M_1^2+M_2^2)}{M_1^2+M_2^2+M_3^2}\right)\,,\\
\\
\tilde{M}_1=\dfrac{\mathrm g}{\tilde{\mathrm g}}
\left(-\dfrac{(\tilde{\g}_2^2+\tilde{\g}_3^2)\g_2\g_3M_1}{\tilde{\g}_2\tilde{\g}_3}
+\dfrac{\g_1\tilde{\g}_2\g_3 M_2}{\tilde{\g}_3}+\dfrac{\g_1\g_2\tilde{\g}_3M_3}{ \tilde{\g}_2 }\right)\,,\\
\\
\tilde{M}_2=\dfrac{\mathrm g}{\tilde{\mathrm g}}
\left(\phantom{-}\dfrac{\tilde{\g}_1\g_2\g_3M_1}{\tilde{\g}_3}
-\dfrac{(\tilde{\g}_1^2+\tilde{\g}_3^2)\g_1\g_3M_2}{\tilde{\g}_1\tilde{\g}_3}
+\dfrac{\g_1\g_2\tilde{\g}_3M_3}{\tilde{\g}_1}\right)\,,\\
\\
\tilde{M}_3=\dfrac{\mathrm g}{\tilde{\mathrm g}}
\left(\phantom{-}
\dfrac{\tilde{\g}_1\g_2\g_3M_1}{\tilde{\g}_2}+\dfrac{\g_1\tilde{\g}_2\g_3M_2}{\tilde{\g}_1}
-\dfrac{(\tilde{\g}_1^2+\tilde{\g}_2^2)\g_1\g_2M_3}{\tilde{\g}_1\tilde{\g}_2}
\right)\,,
\end{array}
\eq
where $\tilde{\mathrm g}$ is the conformal factor (\ref{g-fun}) in $\tilde{\g}$-variables
\[
\tilde{\mathrm g}=\frac{1}{\sqrt{1-d(\tilde{\g},\mathbf A\tilde{\g})}}\equiv
\frac{1}{\sqrt{1-d(a_1\tilde{\g}_1^2+a_2\tilde{\g}_2^2+a_3\tilde{\g}_3^2)}}\,,
\]
which can be also rewritten as a function on the original conformal factor and integrals of motion
\[
\tilde{\mathrm g}^2=\dfrac{1}{\mathrm g^2}\left(\dfrac{dH_1-H_2}{(a_1d-1)(a_2d-1)(a_3d-1)H_2}\right)\,.
\]
This transformation of variables has the following properties:
\begin{prop}
If $(\g,M)=0$ then transformation  (\ref{chap-keyless}) preserves the form of the Poisson bivector $\Pi$ (\ref{chap-poi}), the form of  the  bracket (\ref{chap-br}) and the form of integrals of motion, i.e.
 \[C_1=(\g,\g)=(\tilde{\g},\tilde{\g})\,,\qquad C_2=(\g,M)=(\tilde{\g},\tilde{M})=0\,,\]
 and
\[
H_1=(\mathbf A_\g M,M)=(\mathbf A_{\tilde{\g}} \tilde{M},\tilde{M})\,,\qquad H_2=(M,M)=(\tilde{M},\tilde{M})\,.
\]
\end{prop}
The proof is a straightforward calculation.

\begin{rem}
At $d=0$ conformal factor  is equal to unity $\mathrm g=1$ and transformation
(\ref{chap-keyless}) preserves canonical Poisson brackets on cotangent bundle to sphere $T^*\mathbb S^2$
\bq\label{e3-poi}
\{M_i,M_j\}=\varepsilon_{ijk}M_k,\qquad
 \{M_i, x_j\}=\varepsilon_{ijk}x_k\,,\qquad
\{ x_i, x_j\}=0\,.
\eq
It also preserves the  form of Casimir functions $(\g,\g)=(\tilde{\g},\tilde{\g})$,  $(\g,M)=(\tilde{\g},\tilde{M})=0$ and integrals of motion
\[
H_1=a_1M_1^2+a_2M_2^2+a_3M_3^2=a_1\tilde{M}_1^2+a_2\tilde{M}_2^2+a_3\tilde{M}_3^2 \,,\qquad H_2=M_1^2+M_2^2+M_3^2=\tilde{M}_1^2+\tilde{M}_2^2+\tilde{M}_3^2\,.
\]
Using  this  hidden symmetry of the level set manifold we can  construct a new integrable system on cotangent bundle $T^*\mathbb S^2$.
\end{rem}

Let $(\tilde{u}_{1},\tilde{u}_2,\tilde{p}_{u_1},\tilde{p}_{u_2})$ are images of coordinates  $(u_{1},u_2,p_{u_1},p_{u_2})$ after transformation (\ref{chap-keyless}).  Suppose that two functions
\[
\lambda_k=4(d\tilde{u}_k-1)(a_1-\tilde{u}_k)(a_2-\tilde{u}_k)(a_3-\tilde{u}_k)\tilde{p}_{u_k}^2
\]
are eigenvalues of the recursion operator $N=\Pi'\Pi^{-1}$, where
\bq\label{chap-br2}
\Pi'=\left(
     \begin{array}{cccc}
       0 &0 & \lambda_1 & 0 \\
       0 & 0 & 0 & \lambda_2 \\
       -\lambda_1 & 0 & 0 & 0 \\
       0 & -\lambda_2 & 0 & 0 \\
     \end{array}
   \right)
\eq
is the Poisson bivector  in $(\tilde{u}_{1},\tilde{u}_2,\tilde{p}_{u_1},\tilde{p}_{u_2})$ variables.  The Nijenhuis torsion of $N$ vanishes as a consequence of the compatibility between $\Pi'$ and $\Pi$ (\ref{can-p}).

So, we have a Poisson-Nijenhuis manifold (partial case of bi-Hamiltonian manifolds) endowed with a pair of compatible non-degenerate Poisson brackets. On such manifold functions we can determine functions  $J_m=1/2m\mbox{tr}N$ satisfying the Lenard relations
\[
\Pi'\mathrm dJ_i=\Pi\mathrm d J_{i+1}\,,
\]
which are in involution with respect to both Poisson brackets.

In our case first such  integral of motion is equal to
\bq\label{tH1}
\tilde{H}=\lambda_1+\lambda_2=
(\tilde{\omega},M)=(\tilde{\mathbf A}'_\g M,M)\,,\qquad \tilde{\mathbf A}'_\g=\dfrac{{\mathbf A}-(\g,\mathbf A\g)\mathbf E}{2}+d\mathrm g^2\mathbf B_\g\,,
\eq
where  $\mathbf E$ is a unit matrix and
\[
\mathbf B_\g= \left(\begin{array}{ccc}
 (a_1-a_3)(a_1-a_2)x_1^2&0&0\\
 0&(a_2-a_3)(a_2-a_1)x_2^2&0\\
 0&0&(a_3-a_1)(a_3-a_2)x_3^2\\
 \end{array}\right)
\]
Second integral of motion is the following polynomial of fourth order in momenta
\ben
\tilde{K}&=&\lambda_1\lambda_2=4\mathrm g^2(a_1x_1M_1+a_2x_2M_2+a_3x_3M_3)^2
\Bigl(\bigl(1-a_1d-\mathrm g^2d^2(a_1-a_3)(a_1-a_2)x_1^2\bigr)M_1^2+
\Bigr.\nn\\
\nn\\
&+&\Bigl.\bigl(1-a_2d-\mathrm g^2d^2(a_2-a_1)(a_2-a_3)x_2^2\bigr)M_2^2
+\bigl(1-a_3d-\mathrm g^2d^2(a_3-a_1)(a_3-a_2)x_3^2\bigr)M_3^2
\Bigr).\nn
\en
or
\bq\label{tK1}
\tilde{K}=4\mathrm g^2(\g,\mathbf AM)^2\Bigl((\mathbf E-d\mathbf A-d^2\mathrm g^2\mathbf B_\g) M,M\Bigr)\,.
\eq
It is easy to prove that functions $\tilde{H}$ and $\tilde{K}$ are in  involution with respect to the Poisson bracket (\ref{chap-br}) and Poisson bracket associated with bivector $\Pi'$ (\ref{chap-br2}).

In this case  new angular velocity of the ball
\[
\tilde{\omega}=\omega+\Delta\omega=\mathbf A_\g M+\dfrac{\mathbf A+(\g,\mathbf A,\g)\mathbf E}{2}M
\]
is the additive deformation of initial angular velocity $\omega=\mathbf A_\g M$, that allows us to say about integrable deformation of the Chaplygin ball at $(\g,M)=0$. It will be interesting to study a physical meaning of the corresponding  nonholonomic model.

\subsection{Integrable Hamiltonian systems on the sphere}
If $d=0$, then  the integrals of motion (\ref{tH1}-\ref{tK1}) are equal to
\[
\tilde{H}=(\g,\g)(M,\mathbf  AM)-(\g,\mathbf  A\g)(M,M)\,,\qquad
 \tilde{K}=(\g,\mathbf AM)^2(M,M)\,.
 \]
These functions are in involution with respect to the canonical Poisson bracket (\ref{e3-poi}) and, therefore,
determine integrable Hamiltonian systems  on cotangent bundle to unit sphere $T^*\mathbb  S^2$.  We can also add some potentials to $\tilde{H}$, for instance,
\[
\tilde{H}'=\tilde{H}+a(\g,\mathbf  A\g)+b (\g,\mathbf  A\g)^3
\]
without loss of integrablity. Integrable geodesic flow associated with $\tilde{H}$ has been found in \cite{ts17v} together with another geodesic flows with integrals
\[
\check{H}=(\g,\mathbf A\g)\hat{H}-(\g,\g)(\g,\mathbf AM)^2\,,\qquad
\check{K}=(\g,\mathbf AM)^2\Bigl((\g,\mathbf A\g)(M,\mathbf AM)- (\g,\mathbf A M)^2 \Bigr)\,.
\]

One more integrable geodesic flow we can obtain directly applying arithmetic of divisors associated another equivalence relation, similar bi-Hamiltonian systems on the plane are discussed in  \cite{ts17}. Indeed, let us consider the free motion on the sphere with integrals
\[
H_1=M_1^2+M_2^2+M_3^2\,,\qquad H_2=-(a_1M_1^2+a_2M_2^2+a_3M_3^2)\,.
\]
In elliptic coordinates  $u_{1,2}$ and momenta $p_{u_{1,2}}$
\ben
\g_i&=&\sqrt{\dfrac{(u_1-a_i)(u_2-a_i)}{(a_j-a_i)(a_k-a_i)}}\,
,\qquad i\neq j\neq k\,,\nn\\
\nn\\
M_i&=&\dfrac{2\varepsilon_{ijk}\g_j\g_k(a_j-a_k)}{u_1-u_2}\,
\Bigl((a_i-u_1)p_{u_1}-(a_i-u_2)p_{u_2}\Bigr)\,.
\nn
\en
these integrals read as
\bq\label{eul-ham}
H_1=\dfrac{\phi_1p_{u_1}^2}{u_1-u_2}+\dfrac{\phi_2p_{u_2}^2}{u_2-u_1}\,,\qquad
H_2= -\dfrac{u_2\phi_1p_{u_1}^2}{u_1-u_2}-\dfrac{u_1\phi_2p_{u_2}^2}{u_2-u_1}\,,
\eq
where
\[
\phi_k=4(a_1-u_k)(a_2-u_k)(a_3-u_k)\,,\qquad k=1,2.
\]
The corresponding separation relations  determine genus one hyperelliptic curve
\[
X:\qquad y^2=f(x)\,,\qquad f(x)=4(a_1-x)(a_2-x)(a_3-x)(H_1x+H_2)\,.
\]
Here $x=u_{1,2}$ and $y=\phi_{1,2}p_{u_{1,2}}$ and  Abel's quadratures on this curve have the form
\[
\dfrac{\mathrm du_1}{\sqrt{f(u_1)}}+\dfrac{\mathrm du_2}{\sqrt{f(u_2)}}=0\,,\qquad
\dfrac{u_1\mathrm du_1}{\sqrt{f(u_1)}}+\dfrac{u_2\mathrm du_2}{\sqrt{f(u_2)}}=2\mathrm dt\,.
\]
In order to reduce non-trivial second quadrature to the standard holomorphic form
\bq\label{ab-qv-tau}
\dfrac{\mathrm du_1}{\sqrt{f(u_1)}}+\dfrac{\mathrm du_2}{\sqrt{f(u_2)}}=2\mathrm d\tau
\eq
 we have to change  time $t\to\tau$ and  replace the canonical Poisson bracket $\{.,.\}$ to the bracket  $\{.,.\}_W$
 \bq\label{poi2}
  \{ {u}_i, {p}_{u_j}\}_W= {u}_i\delta_{ij}\,,\qquad  \{ {u}_1, {u}_2\}_W=\{ {p}_{u_1}, {p}_{u_2}\}_W=0\,,
 \eq
 see details in  \cite{ts17}.

Supposing  that variables $\tilde{u}_{1,2}$ and $\tilde{p}_{u_{1,2}}$ satisfy to the same separation relations and  quadratures (\ref{ab-qv-tau}), one gets Abel's differential equation
\bq\label{ab-eq-g1}
\omega(x_1,y_1)+\omega(x_2,y_2)+\omega(x''_1,y''_1)+\omega(x''_2,y''_2)=0
\eq
where $\omega=\mathrm dx/y$ is a holomorphic differential on  an elliptic curve and
\[
x_{1,2}=u_{1,2}\,,\quad y_{1,2}=\phi_{1,2}p_{u_{1,2}}\,,\qquad
x''_{1,2}=\tilde{u}_{1,2}\,,\quad y''_{1,2}=-\tilde{\phi}_{1,2}\tilde{p}_{u_{1,2}}\,.
\]
According \cite{ab} solutions of this equation are the coordinates of  points of intersection $P_ {1,2} = (x_ {1,2}, y_ {1,2} $ and $P''_ {1,2} (x''_ {1,2}, y''_ {1,2} $ of elliptic curve $X$ with the plane curve (conic section)
\[
Y:\qquad y=\mathcal P(x)\,,\qquad   \mathcal P(x)=b_2x^2+b_1x+b_0\,.
\]
The elimination of coefficients $b_i$ leads to determinant
\bq\label{comp-int}
\left|
  \begin{array}{cccc}
     x_1^2 &  x_1 &1 &  y_1 \\
     x_2^2 &  x_2 &1 &  y_2 \\
     x_1'^2 &  x'_1 &1 &  y'_1 \\
     x_2'^2 &  x'_2 & 1 & y'_2 \\
  \end{array}
\right|=0\,,
\eq
as the integral relation corresponding to Abel's differential equation (\ref{ab-eq-g1}).

We can determine coefficients $b_i$ using a standard method \cite{eh16,har77}. By definition four points $P_1,P_2,P_1'',P_2''$
form an intersection divisor of $X$ and $Y$
\[
X\cdot Y=\sum_{P\in X\bigcap Y} \mbox{mult}_P(X,Y)\,P=P_1+P_2+P''_1+P''_2\,.
\]
Note that it is symmetric in $X$ and $Y$, so the result can be considered as an element of Picard groups  Pic$X$ and Pic$Y$, simultaneously.

In the previous Section we consider the addition of divisors in Picard group of hyperelliptic curve $X$
\[
D+D'=D''\,,\qquad  D,D',D''\in \mbox{Pic}^0 X\,,
\]
 where supp$(D)=\{ P_1,P_2\}\cup \{P_\infty\}$ and supp$(D'')=\{ P''_1,P''_2\}\cup \{P_\infty\}$. Now we consider similar  addition of divisors in the second Picard group of conic section $Y$
\[
D+D'=D''\,,\qquad  D,D',D''\in \mbox{Pic}^0 Y\,.
\]
Here supp$D'=\{P_1'\}\cup \{P_\infty\}$ and $P'_1=(x'_1,y'_1)$ is a point on the plane out of $X$. Parabola $Y$ interpolates $P_1,P_2$ and $P'_1$
\[
y_1=\mathcal P(x_1)\,,\qquad y_2=\mathcal P(x_2)\,,\qquad y'_1=\mathcal P(x'_1).
\]
According \cite{ts17},  we  take $P'_1=(0,0)$ and substitute $y=\mathcal P(x)$, where
\bq\label{eul-P}
\mathcal P(x)= x\left(\dfrac{(x-x_2)y_1}{(x_1-x_2)x_1}+\dfrac{(x-x_1)y_2}{(x_2-x_1)x_2}\right)
\eq
due to Lagrange interpolation, into the equation $y^2-f(x)=0$ for  $X$. As a result we obtain Abel's polynomial on $x$
\[
\psi(x)=\mathcal P^2-f(x)=A_2(x-x_1)(x-x_2)(x-x''_1)(x-x''_2)\,,
\]
having four zeroes at the points of intersection.  Equating coefficients of $\psi$   gives abscissas  $x''_1$ and $x''_2$, whereas ordinates are equal to
 \[
 y''_{k}=-\mathcal P(x''_k)\,,\qquad k=1,2.
 \]
In our case
\[
A_2=\dfrac{\bigl(4u_1^2(u_2-u_1)-\phi_1\bigr)\phi_1p_{u_1}^2}{u_1^2(u_1-u_2)^2}
+\dfrac{2\phi_1\phi_2p_{u_1}p_{u_2}}{u_1u_2(u_1-u_2)^2}
+\dfrac{\bigl(4u_2^2(u_1-u_2)-\phi_2\bigr)\phi_2p_{u_2}^2}{u_2^2(u_2-u_1)^2}
\]
 and desired variables $\tilde{u}_{1,2}=x''_{1,2}$ are the roots of the polynomial
 \bq\label{eul-pol}
 \tilde{U}(x)=\dfrac{\psi(x)}{(x-x_1)(x-x_2)}=A_2x^2+A_1x+4a_1a_2a_3\left(
 \dfrac{\phi_1p_{u_1}^2}{u_1(u_2-u_1)}+\dfrac{\phi_2p_{u_2}^2}{u_2(u_1-u_2)}
  \right)\,,
  \eq
 where
  \[
 A_1=\dfrac{\bigl(4u_1^2(a_1+a_2+a_3-u_1)-\phi_1\bigr)\phi_1p_{u_1}^2}{u_1^2(u_1-u_2)}
  +\dfrac{\bigl(4u_2^2(a_1+a_2+a_3-u_2)-\phi_2\bigr)\phi_2p_{u_2}^2}{u_2^2(u_2-u_1)}\,.
  \]
Using equations (\ref{eul-P}-\ref{eul-pol}) we  obtain $x''_{1,2}$ and $y''_{1,2}$ as functions on $x_{1,2}$ and $y_{1,2}$. So, we can explicitly determine auto B\"{a}cklund transformations for the motion on the sphere governed by Hamiltonians $H_{1,2}$ (\ref{eul-ham}).

 \begin{prop}
If  variables $\tilde{u}_{1,2}$ are the roots of polynomial (\ref{eul-pol}) and
\[
\tilde{p}_{u_{k}}=-\dfrac{x}{4(a_1-x)(a_2-x)(a_3-x)}\,
\left.\left(\dfrac{(x-u_2)\phi_1p_{u_1}}{(u_1-u_2)u_1}+\dfrac{(x-u_1)\phi_2p_{u_2}}{(u_2-u_1)u_2}\right)
\right|_{x=\tilde{u}_{k}}\,,
\]
then mapping $(u,p_u)\to (\tilde{u},\tilde{p}_u)$  preserves  the  form of Hamiltonians  $H_{1,2}$ (\ref{eul-ham}), Abel's quadrature (\ref{ab-qv-tau})  and the form of the Poisson bracket $\{.,.\}_W$ (\ref{poi2}), i.e.
\[
\{\tilde{u}_i,\tilde{p}_{u_j}\}_W=\tilde{u}_i\delta_{ij}\,,\qquad  \{\tilde{u}_1,\tilde{u}_2\}_W=\{\tilde{p}_{u_1},\tilde{p}_{u_2}\}_W=0\,.
\]
\end{prop}
The proof is a straightforward calculation.

According \cite{ts17}, in order to get new canonical variables on $T^*\mathbb S^2$ we can use
additional Poisson map
 \[
 \rho:\qquad (u_1,u_2,p_{u_1},p_{u_2})\to (u_1,u_2,u_1p_{u_1},u_2p_{u_2})\,,
 \]
 which  reduces canonical Poisson bracket $\{.,.\}$ (\ref{can-p}) to  bracket $\{.,.\}_W$ (\ref{poi2}). Indeed,
  using the composition of  Poisson mappings  $\rho$ and  $(u,p_u)\to (\tilde{u},\tilde{p}_u)$ we determine variables
 $\hat{u}_{1,2}$, which are the roots of the polynomial
 \bq\label{new-u2}
 \rho(\tilde{U})(x)= \rho(A_2)x^2+\rho(A_1)x+4a_1a_2a_3\left(
 \dfrac{u_1\phi_1p_{u_1}^2}{u_2-u_1}+\dfrac{u_2\phi_2p_{u_2}^2}{u_1-u_2}
  \right)=0
  \eq
The corresponding momenta are equal to
\bq\label{new-p2}
\hat{p}_{u_{k}}=-\dfrac{1}{4(a_1-\hat{u}_{k})(a_2-\hat{u}_{k})(a_3-\hat{u}_{k})}\,
\left(\dfrac{(\hat{u}_{k}-u_2)\phi_1p_{u_1}}{(u_1-u_2)}+\dfrac{(\hat{u}_{k}-u_1)\phi_2p_{u_2}}{(u_2-u_1)}\right)
\eq
The straightforward calculation allows us to prove the following statement.
\begin{prop}
Canonical Poisson bracket
\[
\{u_i,p_{u_j}\}=\delta_{ij}\,,\qquad  \{u_1,u_2\}=\{p_{u_1},p_{u_2}\}=0
\]
 has the same form
\[
\{\hat{u}_i,\hat{p}_{u_j}\}=\delta_{ij}\,,\qquad  \{\hat{u}_1,\hat{u}_2\}=\{\hat{p}_{u_1},\hat{p}_{u_2}\}=0
\]
in variables  $\hat{u}_{1,2}$ and  $\hat{p}_{u_{1,2}}$  (\ref{new-u2},\ref{new-p2}).
\end{prop}
Let $(\hat{u}_{1},\hat{u}_2,\hat{p}_{u_1},\hat{p}_{u_2})$ are images of coordinates  $(u_{1},u_2,p_{u_1},p_{u_2})$ after transformation (\ref{new-u2},\ref{new-p2}).  Suppose that two functions
\[
\nu_k=4\hat{u}_k(a_1-\hat{u}_k)(a_2-\hat{u}_k)(a_3-\hat{u}_k)\hat{p}_{u_k}^2
\]
are eigenvalues of the recursion operator $N=\Pi'\Pi^{-1}$, where
\[
\Pi'=\left(
     \begin{array}{cccc}
       0 &0 & \nu_1 & 0 \\
       0 & 0 & 0 & \nu_2 \\
       -\nu_1 & 0 & 0 & 0 \\
       0 & -\nu_2 & 0 & 0 \\
     \end{array}
   \right)
\]
is the Poisson bivector  in $(\hat{u}_{1},\hat{u}_2,\hat{p}_{u_1},\hat{p}_{u_2})$ variables.  The Nijenhuis torsion of $N$ vanishes as a consequence of the compatibility between $\Pi'$ and $\Pi$ and, therefore, the following functions
\[
\hat{H}=\frac{\nu_1+\nu_2}{2}=\sum_{i,j=1}^2 \hat{\mathrm g}_{ij} p_{u_i}p_{u_j}\,,
\]
where
\[
 \hat{\mathrm g}=\left(
  \begin{array}{cc}
    \dfrac{\phi_1u_1\bigl( 2u_1+u_2-u_1 u_2( a_1^{-1}+a_2^{-1}+a_3^{-1} )\bigr)}{u_1-u_2} & 0 \\
    0 & \dfrac{\phi_2u_2\bigl( u_1+2u_2-u_1 u_2( a_1^{-1}+a_2^{-1}+a_3^{-1} )\bigr)}{u_2-u_1} \\
  \end{array}
\right)
\]
and
\[
\hat{K}=\dfrac{\phi_1\phi_2}{a_1a_2a_3}\left(\dfrac{u_1^2p_{u_1}-u_2^2p_{u_2}}{u_1-u_2}\right)^2
\dfrac{\phi_1u_1p_{u_1}^2-\phi_2u_2p_{u_2}^2}{u_1-u_2}\,.
\]
are in the involution with respect to the canonical Poisson bracket on $T^*\mathbb S^2$.

In terms of original  variables $\g$ and $M$ functions $\hat{H}$ and $\hat{K}$ have more complicated form.
\begin{prop}
If $(\g,\g)=1$ , $(\g,M)=0$ and $\mathbf A$ is the nonsingular diagonal  matrix  then functions
\[
\hat{H}=\alpha (M,\mathbf AM)-\Bigl((\g,\mathbf A^\vee \g)- \alpha (\g,\mathbf A\g)+\alpha\mbox{\rm tr}\mathbf A\Bigr)\cdot (M,M)\,,
\]
where
\[
\alpha=(\g,\mathbf A^\vee \g)\mbox{\rm tr}\mathbf A^{-1}+2 (\g,\mathbf A\g)-2\mbox{\rm tr}\mathbf A\,,
\]
and
\ben
\hat{K}&=&\Bigl((\g,\mathbf A^\vee \g)\cdot(\g,\mathbf A M)-\bigl((\g,\mathbf A\g)-\mbox{\rm tr}\mathbf A\bigr)\cdot(\g,\mathbf A^\vee M)\Bigr)^2\nn\\
\nn\\
&\times&\Bigl(  (M,\mathbf AM) +\bigl((\g,\mathbf A\g)-\mbox{\rm tr}\mathbf A\bigr)\cdot(M,M)  \Bigr)\,,
\en
are in involution with respect  to the canonical Poisson bracket (\ref{e3-poi}). Here  $\mathbf A^\vee=\mathbf A^{-1}\cdot\mbox{\rm det}\mathbf A$ is the cofactor matrix and $\mbox{\rm tr}\mathbf A$ is a trace of matric $\mathbf A$.
\end{prop}
The proof is a straightforward calculation.

Construction of geodesic flows on Riemannian manifolds is a classical object   \cite{ bj04}. A particular place among them is occupied by integrable geodesic flows. We obtain  integrable geodesic flow on the sphere $\mathbb S^2$ with quartic in momenta  integral of motion which is absent in the list of known integrable systems on the sphere.

The work was supported by the Russian Science Foundation (project  15-11-30007).

\end{document}